\documentclass[showpacs,preprintnumbers,eqsecnum,twocolumn,prd]{revtex4}  

\usepackage{graphicx}
\usepackage{amssymb}
\usepackage{subfigure}

\def\beq{\begin{equation}}
\def\eeq{\end{equation}}

\begin{document}

\title{Scalar field inflation and Shan-Chen fluid models}

\author{Donato Bini}
  \affiliation{
Istituto per le Applicazioni del Calcolo ``M. Picone,'' CNR, I-00185 Rome, Italy\\
ICRA, ``Sapienza" University of Rome, I-00185 Rome, Italy\\
INFN, Sezione di Napoli, I--80126 Naples, Italy
}

\author{Andrea Geralico}
  \affiliation{Istituto per le Applicazioni del Calcolo ``M. Picone,'' CNR, I-00185 Rome, Italy\\
ICRA, ``Sapienza" University of Rome, I-00185 Rome, Italy
}

\author{Daniele Gregoris}
  \affiliation{Department of Physics, Stockholm University, 106 91 Stockholm, Sweden\\
Max-Planck-Institut f\"ur Gravitationsphysik (Albert-Einstein-Institut) Am M\"uhlenberg 1, DE-14476 Potsdam, Germany
}

\author{Sauro Succi}
  \affiliation{Istituto per le Applicazioni del Calcolo ``M. Picone,'' CNR, I-00185 Rome, Italy
	}

\begin{abstract}
A scalar field equivalent to a non-ideal \lq\lq dark energy fluid" obeying a Shan-Chen-like equation of state is used as the background source of a flat Friedmann-Robertson-Walker cosmological spacetime to describe the inflationary epoch of our universe.
Within the slow-roll approximation, a number of interesting features are presented, including 
the possibility to fulfill current observational constraints as well as a graceful exit mechanism from the inflationary epoch. 
\end{abstract}


\maketitle

\section{Introduction}

According to the standard inflationary cosmology, the early universe underwent a very short period of rapid accelerated expansion, which was originally assumed as a mechanism to solve several puzzles of the Big Bang theory, e.g., the flatness and horizon problems \cite{guth,sato,starob,linde1,stein,linde2}.
Inflation also provides plausible scenarios for the origin of the large scale structure of the universe, as well as the formation of anisotropies in the cosmic microwave background radiation.
We refer, e.g., to Refs. \cite{linde3,liddle,lyth} for a detailed review on the different kinds of inflationary models developed so far, including recent attempts to construct consistent models of inflation based on superstring or supergravity models.
In order to achieve inflation, the strong energy condition has to be broken (in general), so that the evolution of the universe at very early times is expected to be driven by a source field with positive energy density and negative pressure.
The simplest way to model a system with such a property is to assume a scalar field.
A variety of scalar field models of inflation have been proposed so far, including quintessence \cite{quint}, k-essence \cite{kess1,kess2}, phantom \cite{phan} and tachyonic scalar fields \cite{tach}.

The evolution of a scalar field in a given gravitational background can be equivalently described by using a hydrodynamic representation (see, e.g., Ref. \cite{lyth}).
In fact, the scalar field mainly acts as a perfect fluid, even if it cannot in general be described by an equation of state relating energy density and pressure in a standard way.
Fluids with negative pressure are indeed widely used in cosmology and string theory to describe \lq\lq exotic" matter and topological defects (associated with phase transitions in a primordial universe).
In many practical cases, they can be represented by perfect fluids with a barotropic equation of state $p=w\rho$, with negative $w$, like fluids of cosmic strings and domain walls \cite{cosm1,cosm2}.
The equivalence between field representation and hydrodynamic representation then holds at a very formal level only, and requires some care to be predictive, especially when fluids with negative pressure are involved, as discussed in Ref. \cite{fabris}. 

The usual way to formulate a scalar field model of inflation is to specify a Lagrangian with prescribed kinetic and potential terms, also taking into account the coupling to gravity and eventually to other fields. 
Alternatively, one can start with a \lq\lq dark energy fluid" obeying a given equation of state as a source of the field equations and study the resulting inflationary dynamics. 
Such an approach has been developed, e.g., in Ref. \cite{chap}, where the inflaton field has been assumed to exhibit the same thermodynamic properties of a fluid obeying a Chaplygin gas-like equation of state. 
In the present paper, we follow the same line of thinking as above to develop an inflationary model based on a dark energy field described by a Shan-Chen-like equation of state \cite{shan-chen,sc_prd}.
This equation of state has the suitable property to support a phase transition between low and high density regimes, both characterized by an ideal gas behavior, i.e., pressure and density change in linear proportion to each other.
It was first introduced by Shan and Chen in the context of lattice kinetic theory \cite{shan-chen}, and recently applied to a cosmological context in Ref. \cite{sc_prd} to model the growth of the dark energy component of the present universe.
In the sequel, we will show that this model satisfies current observational constraints over a wide range of parameters, and also provides a a graceful exit mechanism from the inflationary epoch.

\section{Scalar field inflationary models}

The action associated with a scalar field $\phi$ minimally coupled to gravity is given by
\beq
S=\int d^4x\sqrt{-{\rm g}}\left[\frac{M_{\rm pl}^2}{2}R-{\mathcal L}(X,\phi)\right]\,,
\eeq
where ${\rm g}$ is the determinant of the metric, $R$ is the Ricci scalar, $M_{\rm pl}=(8\pi G)^{-1/2}$ is the reduced Planck mass and  ${\mathcal L}$ is the Lagrangian density, which can be in general an arbitrary function of the field and its kinetic term $X=-\frac12g^{\mu\nu}\partial_\mu\phi\partial_\nu\phi$.

Variation of the action with respect to the metric leads to the energy momentum tensor, which can be cast in the form of a perfect fluid, with energy density and pressure given by \cite{dam_mukh}
\beq
\label{rhoandpgen}
\rho=2X{\mathcal L}_X-{\mathcal L}\,,\qquad
p={\mathcal L}\,,
\eeq
where the subscript $X$ indicates differentiation with respect to $X$.
The evolution of a scalar field in a given gravitational background is equivalently described by the motion of a hydrodynamic fluid with a given equation of state.
The hydrodynamic analogy proves useful even if the pressure cannot be expressed in terms of the density alone, because in general $\phi$ and $X$ are independent, as pointed out in Ref. \cite{dam_mukh}. 

In the context of inflation, several models have been proposed by selecting a particular functional form of the Lagrangian density (see, e.g., Refs. \cite{dam_mukh,garr_mukh,dbi_pap,sahni}).
In the present paper, we discuss a canonical scalar field model, i.e., with ${\mathcal L}= X-V(\phi)$, where $X=\dot\phi^2/2$ is the standard kinetic term and $V$ the potential. 
As usual in inflationary cosmology, we also assume the background gravitational field to be a FRW universe with vanishing spatial curvature, i.e., with metric $ds^2=-dt^2+a^2(dx^2+dy^2+dz^2)$, where $a=a(t)$ is the scale factor.
The Euler-Lagrange equations for this system read as follows 
\begin{eqnarray}
\label{ELeqns}
H^2&=&\frac{1}{3M_{\rm pl}^2}\left(\frac{\dot\phi^2}{2}+V\right)\,,\nonumber\\
0&=&\ddot\phi+3H\dot\phi+V'\,,
\end{eqnarray}
where $H=\dot a/a$ is the Hubble parameter.
Here a overdot indicates derivative with respect to cosmic time $t$, whereas a prime denotes differentiation with respect to the scalar field $\phi$.
The hydrodynamic analog of this formulation consists then in taking a perfect fluid as a source of the Einstein's field equations, with energy density and pressure given by
\beq
\label{rhoandp}
\rho=\frac{\dot\phi^2}{2}+V\,,\qquad
p=\frac{\dot\phi^2}{2}-V\,,
\eeq
so that
\beq
\label{Vdef}
V=\frac12(\rho-p)\,,\qquad
\dot\phi^2=\rho+p\,.
\eeq
The field equations 
\beq
\label{FRWeqs}
H^2=\frac{\rho}{3M_{\rm pl}^2}\,,\qquad
\dot\rho=-3H(\rho +p)\,,
\eeq
are then completely equivalent to Eqs.  (\ref{ELeqns}).

\subsection{Slow-roll inflation}

Let us consider a slow-roll approximation for the (background) scalar field, i.e., the case when the potential energy $V$ dominates over the kinetic energy ${\dot\phi^2}/{2}$, driving a quasi-exponential expansion of the universe.
The following conditions hold \cite{liddle}
\beq
\label{SLconds}
\ddot\phi\ll 3H\dot\phi\,,\qquad
\frac{\dot\phi^2}{2}\ll V\,,
\eeq
implying that
\beq
\label{SLconds2}
V'\approx-3H\dot\phi\,,\quad
H^2\approx\frac{V}{3M_{\rm pl}^2}\,,\quad
p\approx-\rho\,.
\eeq

In this case one defines the slow-roll parameters
\begin{eqnarray}
\label{PSRparams}
\epsilon&=&\frac12M_{\rm pl}^2\left(\frac{V'}{V}\right)^2\,,\nonumber\\
\eta&=&M_{\rm pl}^2\frac{V''}{V}\,,\nonumber\\
\Xi&=&M_{\rm pl}^4\frac{V'V'''}{V^2}\,,
\end{eqnarray}
which should remain small, i.e., $\epsilon\ll1$, $|\eta|\ll1$, $|\Xi|\ll1$.
Note that under these conditions $\dot H/H^2\approx-\epsilon$.

Inflation is commonly characterized by the number $N$ of $e$-folds of the expansion.
It is defined as the natural logarithm of the ratio of the scale factor at the final time $t_e$ to its value at some initial time, i.e.,
\beq
\label{eqN}
N=\int_t^{t_e}H dt=\ln\frac{a_e}{a}\,.
\eeq
This measures the amount of inflation that still has to occur after a time $t$, with $N$ decreasing to $0$ at the end of inflation. 
When slow-roll is violated, i.e., $\epsilon(\phi_e)\approx1$ (or $|\eta(\phi_e)|\approx1$, $|\Xi(\phi_e)|\approx1$), the inflationary phase ends.
As it is well known, the universe should then be reheated, undergoing a transition to the radiative phase of the standard cosmological model.
The stability issue related to scalar perturbations of scalar field inflationary models in the slow-roll regime is briefly recalled in the Appendix.

In the following sections, we will develop a canonical scalar field model of inflation based on a dark energy field described by a suitable non-ideal equation of state supporting phase-transitions.

\section{The Shan-Chen fluid model}

Before discussing the application to inflationary cosmology, we shall review below the basic features of non-ideal fluids obeying a Shan-Chen-like (SC) equation of state \cite{shan-chen,sc_prd}, with special focus on the underlying microphysics and its relevance to cosmological models.

\subsection{Microscopic foundations}

Shan and Chen first used a non-ideal equation of state to model dynamic phase transitions in the context of lattice kinetic theory \cite{shan-chen}. The main motivation was to circumvent the problem of the very small time-steps imposed by the short-range (hard-core) repulsion of atomic potentials in the numerical integration 
of the equations of motion of molecular fluids. 
The distinctive feature of the SC equation of state is the replacement of hard-core repulsive interactions, which are needed to tame unstable density build-up, with a purely attractive force, with the peculiar property of becoming vanishingly small above a given density threshold, so as to prevent the onset of instabilities due to uncontrolled density pile-up.   
Since high-density implies short spatial separation, the SC potential implements a form of effective of ``asymptotic freedom,'' meaning by this that molecules below a certain separation behave basically like free particles. 
Since its inception, the SC method has met with major success for the numerical simulation 
of a broad variety of complex flows with phase-transitions \cite{LB1,LB2}. 

More in detail, with specific reference to {\it lattice} fluids, for which the time-step is fixed by the lattice size, Shan and Chen \cite{shan-chen} proposed a \lq\lq synthetic" repulsion-free potential of the form:
\beq
\label{SCPOT}
V({\mathbf x},{\mathbf x}')=\psi({\mathbf x})G({\mathbf x}-{\mathbf x}')\psi({\mathbf x}')\,, 
\eeq
where the generalized density $\psi({\mathbf x}) = \psi[\rho({\mathbf x})]$ is a local functional 
of the fluid density and $G({\mathbf x}-{\mathbf x}')$ is the Green function of the interaction. 
In the above
\beq
{\mathbf x}'={\mathbf x}+{\mathbf e}_a\,,
\eeq
where ${\mathbf e}_a$ denotes a generic spatial direction in the lattice.
The explicit dependence on time of the various functions defined above has been omitted for notational simplicity. 
For instance, a typical two-dimensional lattice features one rest particle ($|{\mathbf e}_0 |=0$), $4$ nearest-neighbors ($|{\mathbf e}_a | =c_L \Delta t$), and $4$ next-nearest-neighbors  ($|{\mathbf e}_a| =c_L \Delta t \sqrt 2)$, $c_L={\Delta x}/{\Delta t}$ being the lattice \lq\lq light speed." 

Shan and Chen took $G({\mathbf x}-{\mathbf x}')={\mathcal G}$ for 
$|{\mathbf e}_a |<c_L \Delta t \sqrt 2$ and zero elsewhere, where ${\mathcal G}<0$ codes 
for attractive interaction.
The associated force per unit volume of the fluid reads as follows: 
\begin{equation}
\label{SCFORCE}
{\mathbf F}({\mathbf x})=-\psi({\mathbf x}){\mathcal G}\sum_a\psi({\mathbf x}+{\mathbf e}_a){\mathbf e}_a\,,
\end{equation}
which equals $-{\mathcal G} c_s^2 \nabla \psi^2/2$ in the limit $\Delta t \to 0$.
Taylor expanding the above expression yields:
\begin{equation}
\label{Force_approx}
{\mathbf F}({\mathbf x}) = -{\mathcal G} \psi({\mathbf x}) \nabla \psi({\mathbf x}) + O(\Delta t^3)\,,
\end{equation}
where we have taken into account that $\sum_a{\mathbf e}_a^i=0$ and 
$\sum_a{\mathbf e}_a^i{\mathbf e}_a^j= (c_L^2 \Delta t^2/3)\delta^{ij}$.

Higher order terms describe physical properties, such as surface tension, which play 
a crucial role in the dynamics of complex fluids.
It is readily shown that the above force contributes an
excess pressure of the form (in lattice units $\Delta t = \Delta x = c_L =1$)
\begin{equation}
\label{pstar}
\frac{p}{c_s^2}-\rho = \frac{\mathcal G}2 \psi^2(\rho)\,,
\end{equation}
$c_s$ being the sound speed of the ideal fluid, typically $c_s^2=1/3$ in lattice units. 

Note that for attractive interactions, i.e., ${\mathcal G}<0$, the excess pressure is negative, leading
to a Van der Waals-like loop for ${\mathcal G}$ sufficiently negative.   
The functional form $\psi(\rho)$ was chosen in Ref. \cite{shan-chen} in such a way as 
to realize a vapor-liquid coexistence curve, i.e.,
\begin{equation}
\psi(\rho)= \rho_0  \left(1-e^{-\frac{\rho}{\rho_0}}\right)\,,
\end{equation}
where $\rho_0$ is a reference density, above which \lq\lq asymptotic freedom" sets in.
The above expression delivers a phase transition at $\rho_c = \rho_0 \ln2$ and ${\mathcal G}<{\mathcal G}_c=-4$.
In the low density regime ($\rho \ll \rho_0$) $\psi \rightarrow \rho$ and the Shan-Chen equation of state (\ref{pstar}) reduces to $p/c_s^2 = \rho + {\mathcal G} \rho^2/2$, which be unstable in the high density regime. 
Due to the saturation of $\psi$ for $\rho\gg\rho_0$, in the high-density
limit the equation of state delivers instead $p/c_s^2 = \rho + {\mathcal G} \rho_0^2/2$.
An interesting feature of the Shan-Chen equation of state is the fact of supporting
a negative pressure for ${\mathcal G}$ sufficiently below ${\mathcal G}_c$, jointly with a positive $c_s^2$.

Finally, it is worth noting that $\psi({\mathbf x})$ can also be interpreted as a scalar field, interacting 
via gauge quanta, whose propagator is precisely $G({\mathbf x}-{\mathbf x}')$ in Eqs. (\ref{SCPOT})--(\ref{SCFORCE}). 
In the standard version, such gauge quanta do not propagate beyond the first Brillouin 
region, the one associated with the $8$ lattice connections described above.
More recent variants of the model also include longer-range interactions, implementing the
competition between short-range attraction and long-range repulsion, which permits to describe
the dynamics of highly complex fluids such as foams and emulsions  \cite{falcucci}.

\subsection{Application to cosmology}

\begin{figure}
\centering
\subfigure{\includegraphics[scale=0.2]{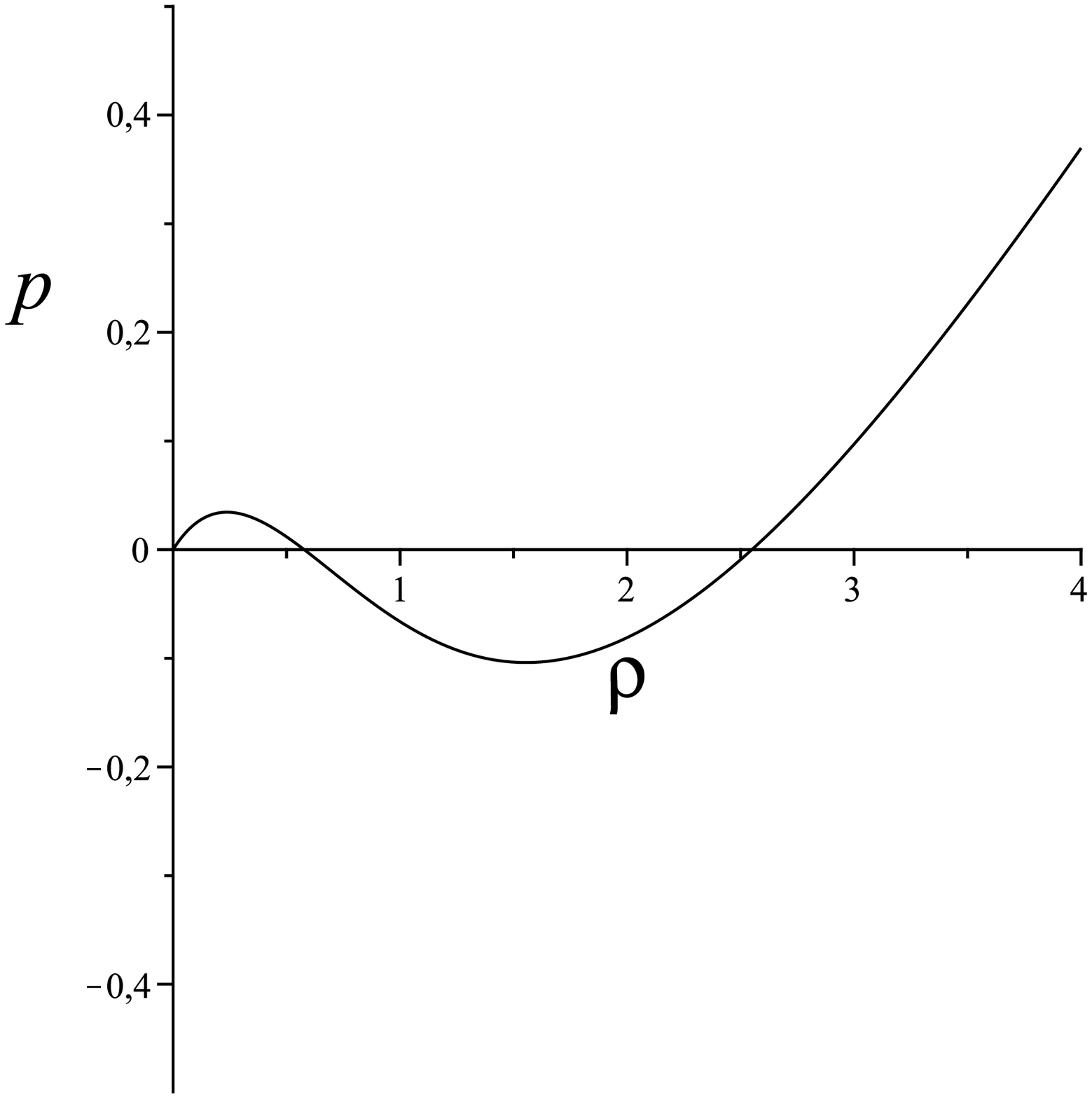}}
\hspace{5mm}
\subfigure{\includegraphics[scale=0.2]{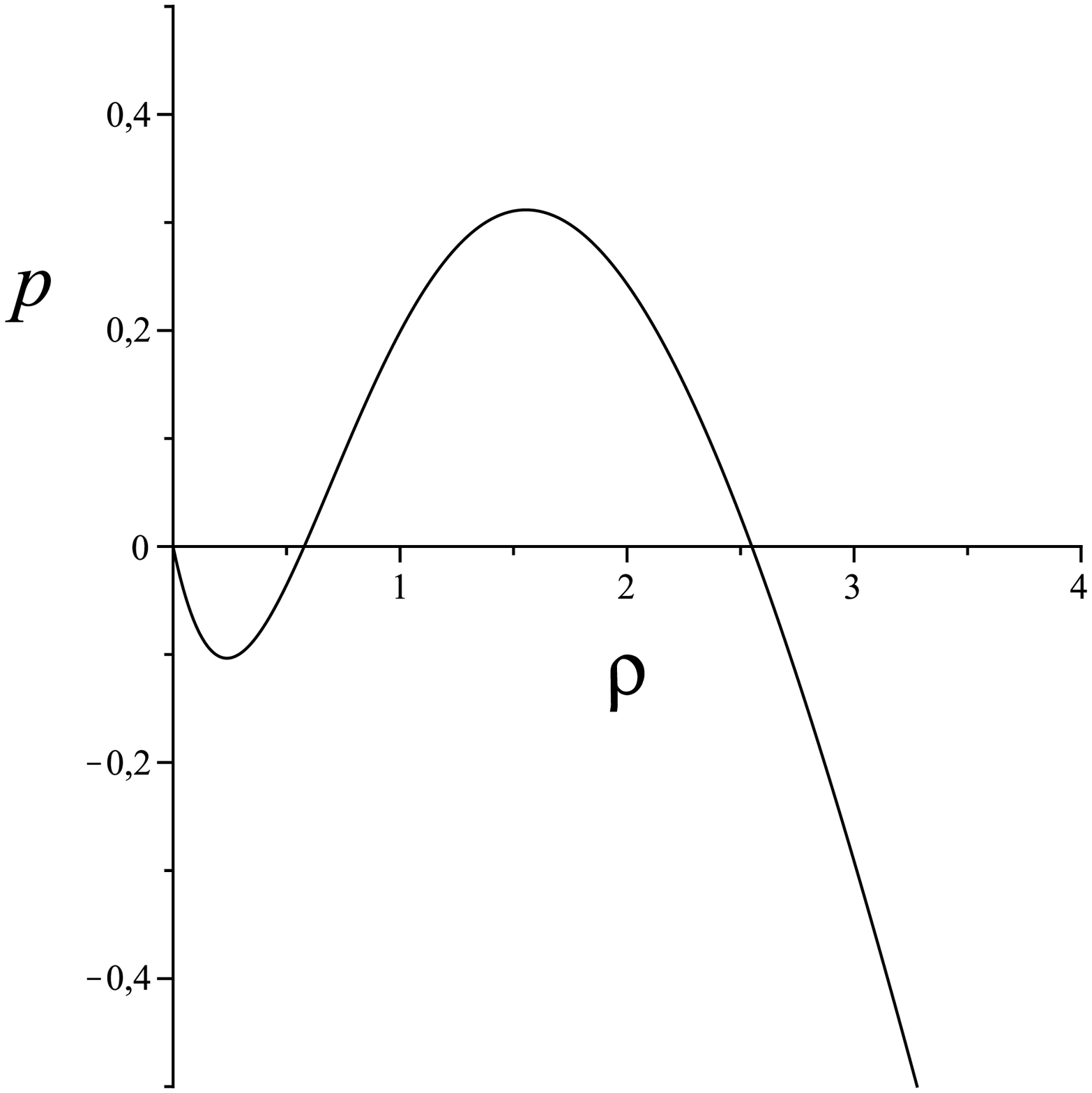}}
\caption{
The behavior of the SC pressure $p$ given by Eq. (\ref{pscdef}) is shown as a function of $\rho$ in units of $\rho_{\rm(crit),0}$ for the choice of parameters $\alpha=1$, $g=-6$ and different values of $w_{\rm (in)}=1/3$ (left) and $w_{\rm (in)}=-1$ (right).  
}
\label{fig:0}
\end{figure}

The SC equation of state is particularly interesting for cosmological applications because of its flexibility, which
permits to represent a broad variety of cosmological fluids, from radiation to dark energy, through a 
smooth variation of its free parameters (see Fig. \ref{fig:0}, for a simple illustration).
Furthermore, it supports phase transitions, so that different epochs in the evolution of the universe 
can be described as a natural consequence of the SC fluid dynamic equations. 

As an example of such flexibility, we have recently presented in Ref. \cite{sc_prd} a new class of cosmological models consisting of a FRW universe with a fluid source, obeying a SC-like equation of state of the form:
\begin{eqnarray}
\label{pscdef}
p&=&w_{\rm (in)}\rho_{\rm (crit),0}  \left[\frac{\rho}{\rho_{\rm (crit),0} }+\frac{g}{2} \psi^2\right]\,,\nonumber\\
\psi&=& 1-e^{-\alpha \frac{\rho}{\rho_{\rm (crit),0}}}\,,
\end{eqnarray}
where $\rho_{\rm (crit),0}=3H_0^2M_{\rm pl}^2$ is the present value of the critical density ($H_0$ denoting the Hubble constant) and the dimensionless quantities $w_{\rm (in)}$, $g \le 0$ and $\alpha\ge0$ are free parameters of the model. 
Note that these three parameters respond to a well-defined physical
interpretation: $w_{\rm (in)}$ describes the nature of matter, ordinary (positive) 
versus exotic (negative). The parameter $g$ measures the strength
of non-ideal interactions within the matter component. 
Finally, $\alpha$ sets the ratio between the actual
critical density and the density above which the excess pressure saturates
to a constant value, a regime sometimes associated with asymptotic-freedom, as 
it corresponds to a vanishing contribution of non-ideal forces 
to the momentum budget of the fluid.

We have shown in Ref. \cite{sc_prd} that, starting from an ordinary equation of state at early times (e.g., satisfying the energy condition typical of a radiation-dominated universe), a cosmological FRW fluid obeying the SC equation of state naturally evolves towards a present-day universe with a suitable dark-energy component, as a consequence of the fluid evolution equations. In the sequel, we shall explore the possibility to develop an inflationary model based on a dark energy field described by a SC-like equation of state of the form (\ref{pscdef}).


\section{Shan-Chen inflation}

Let us turn to the scalar field formulation of inflation briefly recalled in Section II, assuming that the scalar field be described by a fluid satisfying the non-ideal Shan-Chen-like equation of state (\ref{pscdef}).
It proves convenient to introduce the following set of dimensionless variables:
\beq
\xi=\frac{\rho}{\rho_{\rm (crit),0}}\,, \qquad
x=\frac{a}{a_0}\,, \qquad 
\tau=H_0t\,,
\eeq
so that for an expanding universe ($H>0$) Eqs. (\ref{FRWeqs}) reduce to
\beq
\label{FRWeqs2}
\frac{dx}{d\tau }=x\sqrt{\xi}\,,\qquad
\frac{d\xi}{d\tau }=-3\sqrt{\xi}[\xi +w_{\rm (in)}{\mathcal P}(\xi)]\,,
\eeq
where
\beq
\label{mathP}
{\mathcal P}(\xi)=\xi +\frac12 g \left(1-e^{-\alpha \xi }\right)^2\,.
\eeq
Besides the trivial solution $\xi=0=x$, the above system admits as fixed points the solutions of the following equation:
\beq
\label{attractor}
\xi_* +w_{\rm (in)}{\mathcal P}(\xi_*)=0\,,
\eeq
which can be at most two, for fixed values of $\alpha$ and $g$, as discussed in Ref. \cite{sc_prd}.

\begin{figure}
\begin{center}
\includegraphics[scale=0.35]{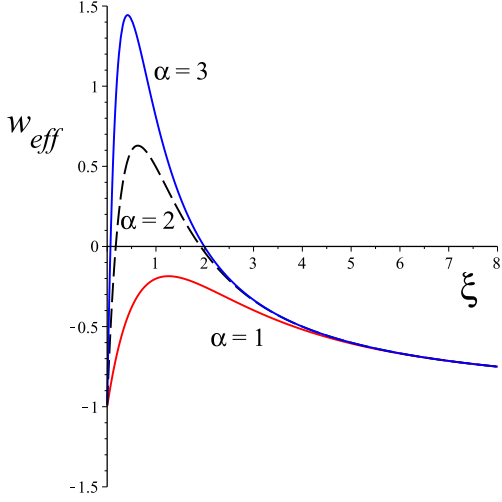}
\end{center}
\caption{
The behavior of the ratio $w_{\rm eff}\equiv p/\rho$ as a function of $\xi$ is shown for the choice of parameters $w_{\rm (in)}=-1$, $g=-4$ and different values of $\alpha=[1,2,3]$.
The dashed curve (with $\alpha=2$) corresponds to the set of parameters used below for the numerical integration of the model equations.
}
\label{fig:1}
\end{figure}

Note that Eq. (\ref{mathP}) reduces to the ideal gas expression ${\mathcal P}(\xi) \sim \xi$  in the
low-density limit $\alpha \xi \to 0$, while in the opposite high-density limit, it delivers
${\mathcal P} \sim \xi + \frac 12 g$, i.e., the non-ideal gas contribution reduces to a 
constant, often associated with vacuum fluctuations.
For more details on the SC thermodynamics, see Ref. \cite{sc_prd}.
Here, we simply note that by changing the three parameters at hand, i.e., $w_{\rm (in)}$, $g$ and $\alpha$, 
the SC equation of state can attain a broad range of values of
cosmological interest for the parameter $w_{\rm eff} \equiv p/\rho$.  
Its behavior as a function of $\xi$ is shown in Fig. \ref{fig:1} for fixed values of $w_{\rm (in)}$ and $g$ and different values of $\alpha$.
Changing the value of $w_{\rm (in)}$ permits to obtain different low-density asymptotic regimes.
Increasing the value of $|g|$ implies that the relative maxima attain larger values. 
Direct inspection of this plot shows that there exist values of $\alpha$ such that the SC equation of state undergoes a transition from exotic matter ($w_{\rm eff}<0$) to ordinary matter ($w_{\rm eff}>0$).


According to the scalar field description discussed above, Eq. (\ref{Vdef}) implies that the 
potential associated with the SC energy density and pressure is given by:
\beq
\label{potdef}
\frac{V}{\rho_{\rm (crit),0}}=\frac12\left(\xi-w_{\rm (in)}{\mathcal P}\right)\,,
\eeq
and the evolution equation for the dimensionless scalar field $\widetilde\phi=\phi/M_{\rm pl}$ reads as:
\beq
\label{eqphi}
\frac{d{\widetilde\phi}}{d\tau}=\sqrt{3(\xi+w_{\rm (in)}{\mathcal P})}\,.
\eeq
The slow-roll parameters (\ref{PSRparams}) turn out to be
\begin{eqnarray}
\label{PSRparams2}
\epsilon&=&\frac{3}{2}\frac{\xi(\xi+w_{\rm (in)}{\mathcal P})}{(\xi-w_{\rm (in)}{\mathcal P})^2}[1-w_{\rm (in)}+g\alpha w_{\rm (in)}\psi(\psi-1)]^2\,,\nonumber\\
\eta&=&-\frac{3}{4(\xi-w_{\rm (in)}{\mathcal P})}\sum_{k=0}^{4}A_k\psi^k\,,\nonumber\\
\Xi&=&\frac{9}{2}\frac{\xi(\xi+w_{\rm (in)}{\mathcal P})[1-w_{\rm (in)}+g\alpha w_{\rm (in)}\psi(\psi-1)]}{(\xi-w_{\rm (in)}{\mathcal P})^2}\nonumber\\
&&\times\sum_{k=0}^{4}B_k\psi^k\,,
\end{eqnarray}
where $\psi=1-e^{-\alpha\xi}$ is defined in Eq. (\ref{pscdef}), and 
\begin{eqnarray}
A_0&=&4\xi(1+w_{\rm (in)})(\xi\alpha^2 gw_{\rm (in)}+w_{\rm (in)}-1)\,,\nonumber\\
A_1&=&2\xi gw_{\rm (in)}\alpha[3w_{\rm (in)}+1-6\alpha\xi(w_{\rm (in)}+1)]\,,\nonumber\\
A_2&=&gw_{\rm (in)}\left[8\alpha^2\xi^2(1+w_{\rm (in)})+w_{\rm (in)}-1\right.\nonumber\\
&&\left.
+2\alpha\xi\left(2\alpha gw_{\rm (in)}-3w_{\rm (in)}-1\right)\right]\,,\nonumber\\
A_3&=&-\alpha g^2w_{\rm (in)}^2(10\alpha\xi-1)\,,\nonumber\\
A_4&=&\alpha g^2w_{\rm (in)}^2(6\alpha\xi-1)\,.
\end{eqnarray}
and
\begin{eqnarray}
B_0&=&2(1-w_{\rm (in)}^2)-g\alpha^2w_{\rm (in)}\xi(5+7w_{\rm (in)})\nonumber\\
&&+6(1+w_{\rm (in)})g\alpha^3w_{\rm (in)}\xi^2\,,\nonumber\\
B_1&=&-4g\alpha w_{\rm (in)}^2+g\alpha^2w_{\rm (in)}\xi(21w_{\rm (in)}-4g\alpha w_{\rm (in)}+15)\nonumber\\
&&-14(1+w_{\rm (in)})g\alpha^3w_{\rm (in)}\xi^2\,,\nonumber\\
B_2&=&4g\alpha w_{\rm (in)}^2-\frac72g^2\alpha^2w_{\rm (in)}^2+8(1+w_{\rm (in)})g\alpha^3w_{\rm (in)}\xi^2\nonumber\\
&&-g\alpha^2w_{\rm (in)}\xi(10-19g\alpha w_{\rm (in)}+14w_{\rm (in)})\,,\nonumber\\
B_3&=&g^2\alpha^2w_{\rm (in)}^2\left(\frac{17}{2}-27\alpha\xi\right)\,,\nonumber\\
B_4&=&g^2\alpha^2w_{\rm (in)}^2(12\alpha\xi-5)\,.
\end{eqnarray}

Next, we study the behaviors of $V$, $\epsilon$, $\eta$ and $\Xi$ as functions of $\xi$ for fixed values of the parameters.
Approaching the initial singularity, $\xi$ goes to infinity or to a value of equilibrium ($\xi_*$).
In the former case we obtain: 
\beq
\lim_{\xi\to\infty}\frac{V}{\rho_{\rm (crit),0}}=-\frac14w_{\rm (in)}g+\frac{\xi}{2}(1-w_{\rm (in)})\,,
\eeq
and
\begin{eqnarray}
\lim_{\xi\to\infty}\epsilon&=&\frac32(1+w_{\rm (in)})\,,\qquad
\lim_{\xi\to\infty}\eta=3(1+w_{\rm (in)})\,,\nonumber\\
\lim_{\xi\to\infty}\Xi&=&9(1+w_{\rm (in)})^2\,.
\end{eqnarray}
When $\xi\to\xi_*$ we have instead that $\epsilon\to0$ and $\Xi\to0$, whereas $\eta\to\eta(\xi_*)\equiv\eta_*$, with $\xi_*$ satisfying Eq. (\ref{attractor}).
The fixed points $\xi_*$ are inflationary attractors, corresponding to an exponential inflation, i.e.,
\beq
\frac{H_*^2}{H_0^2}=\xi_*\,,\qquad
x=e^{\sqrt{\xi_*}\tau}\,,
\eeq
as follows from Eqs. (\ref{FRWeqs}).

For completeness, at $\xi=0$ the potential behaves as
\beq
\lim_{\xi\to0}\frac{V}{\rho_{\rm (crit),0}}=\frac{\xi}{2}(1-w_{\rm (in)})\,,
\eeq
whereas the slow-roll parameters $\epsilon$, $\eta$ and $\Xi$ all approach the same values as in the limit $\xi\to\infty$ shown above.

\subsection{Observational constraints}

In the context of slow-roll inflation, the parameters (\ref{PSRparams}) plus the amplitude of the potential are sufficient to determine the observable quantities. 
For instance, the ratio of tensor to scalar perturbations, the scalar spectral index and its running exponent are given by:
\begin{eqnarray}
r&=&16\epsilon\,,\nonumber\\
n_{s}&=&1-6\epsilon+2\eta\,,\nonumber\\
\alpha_{s}&\equiv&\frac{dn_{s}}{d \ln k}=16\epsilon\eta-24\epsilon^2-2\Xi\,,
\end{eqnarray}
respectively.
Recent cosmic microwave background data from Planck \cite{planck}, combined with the large angle polarization data from the Wilkinson Microwave Anisotropy Probe (WMAP) \cite{wmap} impose strong bounds on these parameters: $r<0.11$ and $n_{s}=0.9603\pm0.0073$ at $95\%$ confidence level. Furthermore, Planck data do not indicate any statistically significant running of the spectral index, i.e., $\alpha_{s}=-0.0134\pm0.0090$.
Finally, the combination of the very recent results of BICEP2 \cite{bicep2} with the other experiments favors a value of the tensor-to-scalar ratio $r$ between $0.13$ and $0.25$ (with $0.2$ preferred).

\subsection{Results}

In order to study the implications of observational constraints on the SC potential and the free parameters of the model, we have first to check that the slow-roll conditions are fulfilled.
Inflation ends if/when $\epsilon\approx1$ (or $|\eta|\approx1$, $|\Xi|\approx1$).
This condition fixes the value of the field $\widetilde\phi_e$ at the end of inflation.
Next, the potential must allow for a sufficient number of inflationary $e-$folds between horizon crossing for observable scales and the end of inflation, which is typically about 60.
Requiring that $N_i\approx60$ at the beginning of inflation provides a condition on the initial value $\widetilde\phi_i$ of the field in order to obtain a sufficient number of $e-$folds, which is assumed to be $N_e=0$ at the end of inflation (with $\widetilde\phi_e=0$ too).

It is convenient to express the evolution of the scalar field $\widetilde\phi$ and the number of $e-$folds $N$ in terms of $\xi$ as follows: 
\begin{eqnarray}
\label{eqsNphi}
\frac{d\widetilde\phi}{d\xi}&=& -\frac1{\sqrt{3\xi(\xi+w_{\rm (in)}{\mathcal P})}}\,,\nonumber\\
\frac{dN}{d\xi}&=& \frac1{3(\xi+w_{\rm (in)}{\mathcal P})}\,,
\end{eqnarray}
from Eqs. (\ref{eqN}) and (\ref{eqphi}).
Since in the limit of  both high and low density the SC equation of state reduces to $p\sim w_{\rm (in)}\rho$, we assume hereafter $w_{\rm (in)}=-1$, to match the inflationary condition also in these asymptotic regimes. 
This choice allows for an analytical study of the model equations.

In fact, the field equations (\ref{FRWeqs2}) reduce to
\beq
\label{FRWeqs3}
\frac{dx}{d\tau }=x\sqrt{\xi}\,,\qquad
\frac{d\xi}{d\tau }=\frac32g\sqrt{\xi}(1-e^{-\beta})^2\,.
\eeq
The latter equation can be also written as
\beq
\frac{d\beta}{dx }=-\frac1{\sigma^2x}(1-e^{-\beta})^2\,,
\eeq
in terms of $\beta=\alpha\xi$ and the new parameter $\sigma^2=2/(3\alpha|g|)$ which summarizes the whole dependence on the parameters of the model.
The solution is 
\beq
\beta=\ln\left(1+\frac1y\right)\,,\qquad
y=W(x^{1/\sigma^2}e^{C/\sigma^2})\,,
\eeq
where $W(z)$ denotes the Lambert $W$ function (see, e.g., Ref. \cite{corless} for its definition and main properties) and $C$ is an integration constant.
Upon inverting we obtain 
\beq
x=e^{-\sigma^2\beta}(1-e^{-\beta})^{-\sigma^2}e^{-C+\sigma^2/(1-e^{-\beta})}\,.
\eeq

Eqs. (\ref{eqsNphi}) then become
\begin{eqnarray}
\label{eqsNphi2}
\frac{d\widetilde\phi}{d\beta}&=& -\frac{\sigma}{\sqrt{\beta}(1-e^{-\beta})}\,,\nonumber\\
\frac{dN}{d\beta}&=& \frac{\sigma^2}{(1-e^{-\beta})^2}\,.
\end{eqnarray}
The solution for $N$ is straightforward 
\beq
\label{solN}
N-N_0= \sigma^2\left[\beta-\frac1{1-e^{-\beta}}+\ln(1-e^{-\beta})\right]\,.
\eeq
The solution for $\widetilde\phi$ instead can only be given formally, i.e.,
\beq
\label{solphi}
\widetilde\phi-\widetilde\phi_0=-\sigma\sum_{k=0}^{\infty}\sqrt{\frac{\pi}{k}}{\rm erf}(\sqrt{k\beta})\,,
\eeq
where ${\rm erf}(z)$ denotes the error function and the term $k=0$ is taken as a limit.  
We have used here the geometric series representation for $(1-e^{-\beta})^{-1}$, whose convergence is ensured for $\beta\gg 1$. 
The quantities $N_0$ and $\widetilde\phi_0$ in Eqs. (\ref{solN}) and (\ref{solphi}) are integration constants.

\begin{figure}
\begin{center}
\includegraphics[scale=0.32]{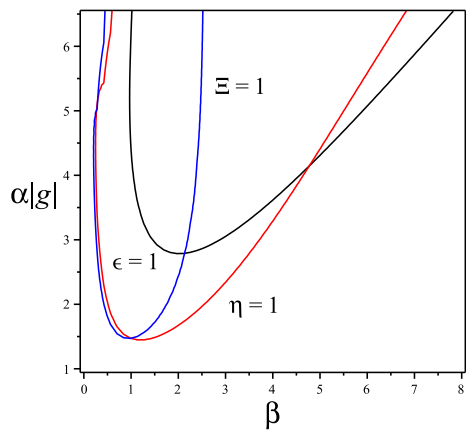}
\end{center}
\caption{
The curves $\epsilon=1$, $\eta=1$ and $\Xi=1$ are implicitly plotted as functions of $\alpha|g|$ and $\beta$ for $w_{\rm (in)}=-1$.
The slow-roll regime turns out to be valid in the region below each curve.
}
\label{fig:2}
\end{figure}

The potential (\ref{potdef}) in this case becomes
\beq
\frac{\alpha V}{\rho_{\rm (crit),0}}=\beta-\frac1{6\sigma^2}\psi^2\,,
\eeq
and the slow-roll parameters (\ref{PSRparams2}) take the following compact form:
\begin{eqnarray}
\epsilon&=&
2\psi^2\beta\frac{(3\sigma^2-\psi+\psi^2)^2}{\sigma^2(-6\beta\sigma^2+\psi^2)^2}\,,\nonumber\\
\eta&=&\frac{\psi}{\sigma^2(-6\beta\sigma^2+\psi^2)}[(6\beta-1)\psi^3+(1-10\beta)\psi^2\nonumber\\
&&
+(6\beta\sigma^2+4\beta-3\sigma^2)\psi-6\beta\sigma^2]\,,\nonumber\\
\Xi&=&-\frac{2(1-\psi)(3\sigma^2-\psi+\psi^2)\psi^2\beta}{\sigma^4(-6\beta\sigma^2+\psi^2)^2}
[2(12\beta-5)\psi^3\nonumber\\
&&
+(7-30\beta)\psi^2+4(3\beta\sigma^2+2\beta-3\sigma^2)\psi-6\beta\sigma^2]\,,\nonumber\\
\end{eqnarray}
where $\psi=1-e^{-\beta}$ as from Eq. (\ref{pscdef}).
Fig. \ref{fig:2} shows the boundary of the region in the parameter space $(\beta,\alpha|g|)$ where the slow-roll regime holds.
The asymptotic behavior of the potential for large values of $\beta$ is then $V\sim\beta$.
In this limit, from Eq. (\ref{eqsNphi2}), one also has $\widetilde\phi\sim\sqrt{\beta}$, implying that $V\sim\widetilde\phi^2$.


\begin{figure}
\centering
\subfigure[]{\includegraphics[scale=0.2]{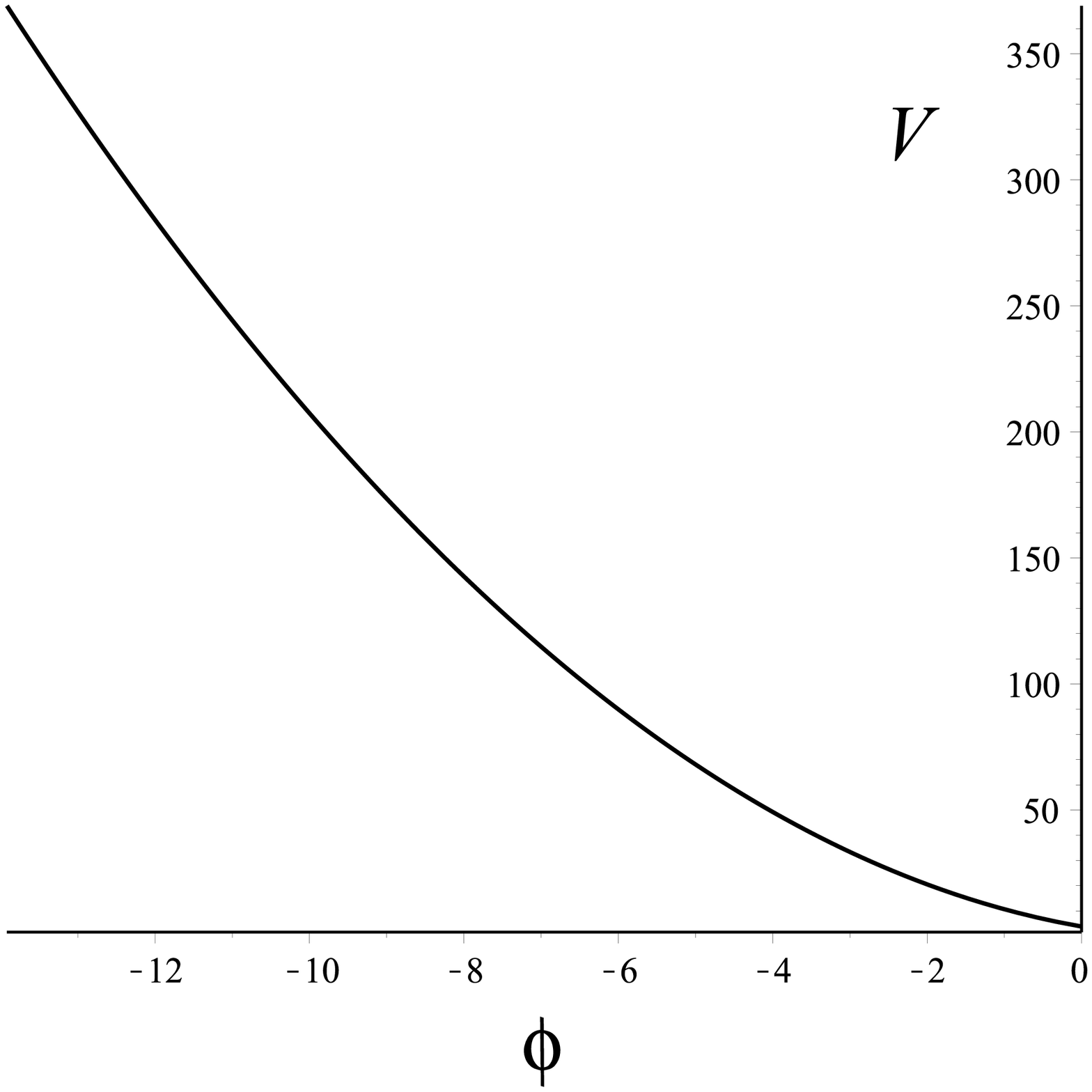}}
\hspace{5mm}
\subfigure[]{\includegraphics[scale=0.2]{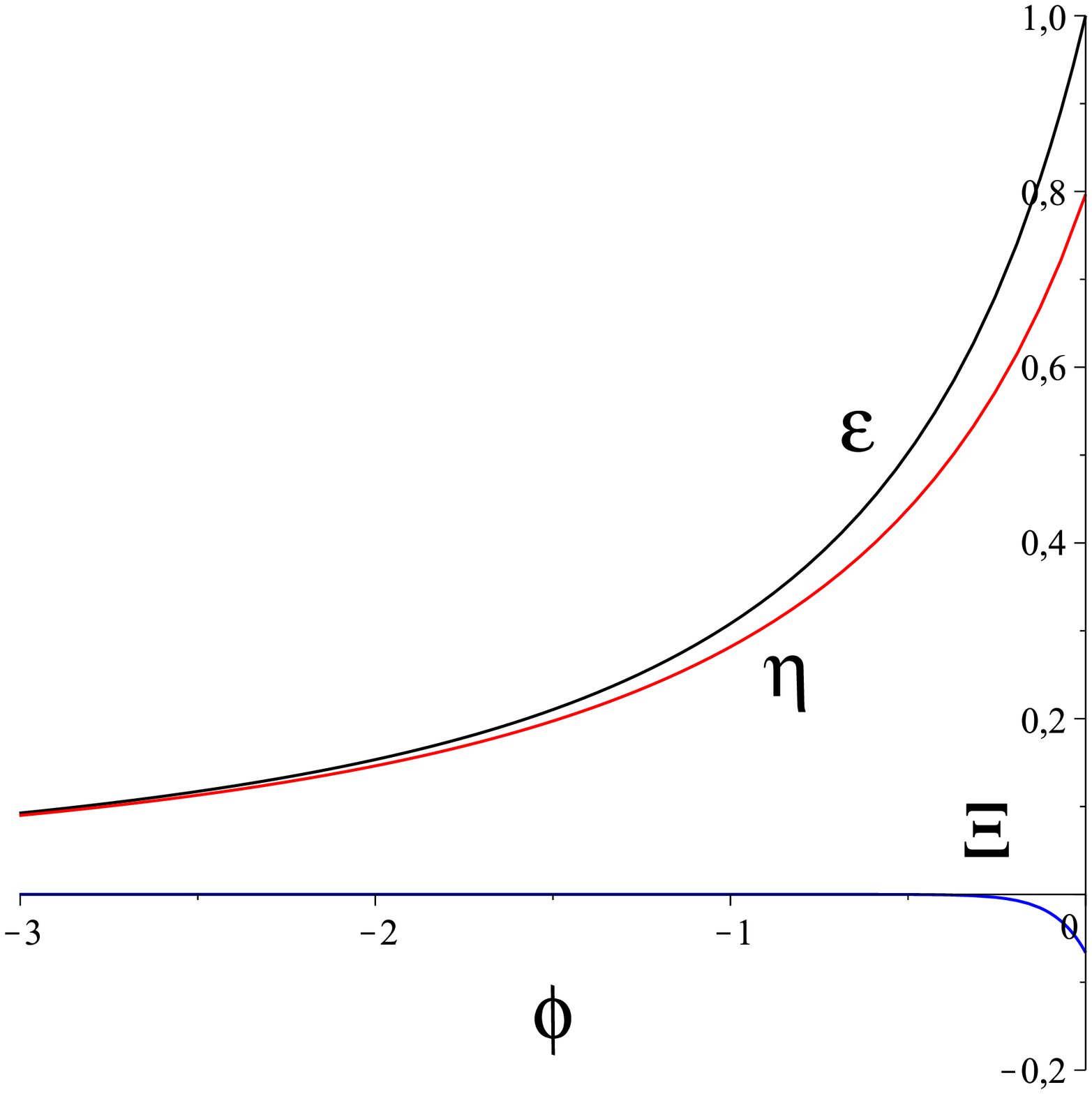}}\\[1cm]
\subfigure[]{\includegraphics[scale=0.2]{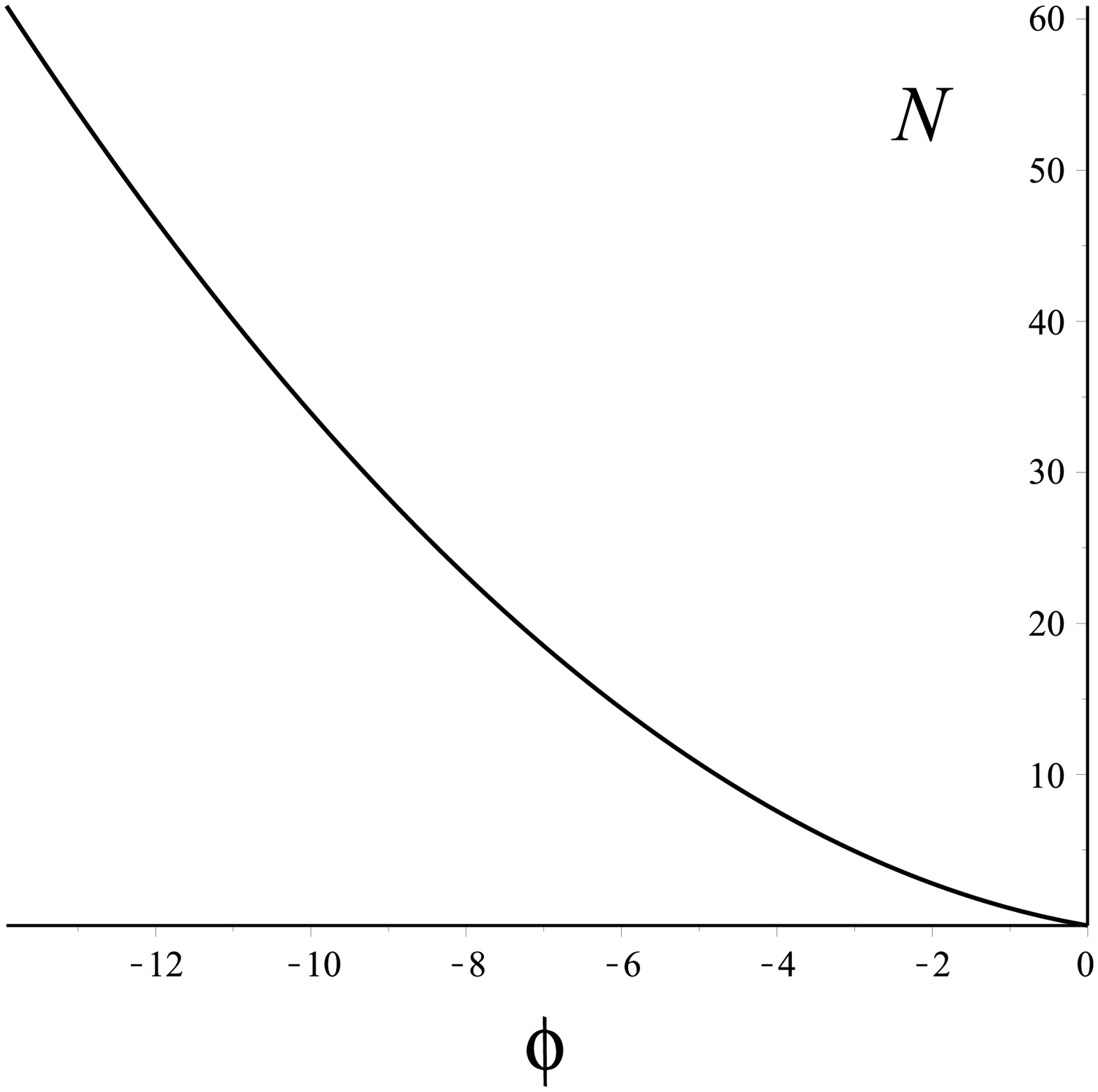}}
\hspace{5mm}
\subfigure[]{\includegraphics[scale=0.2]{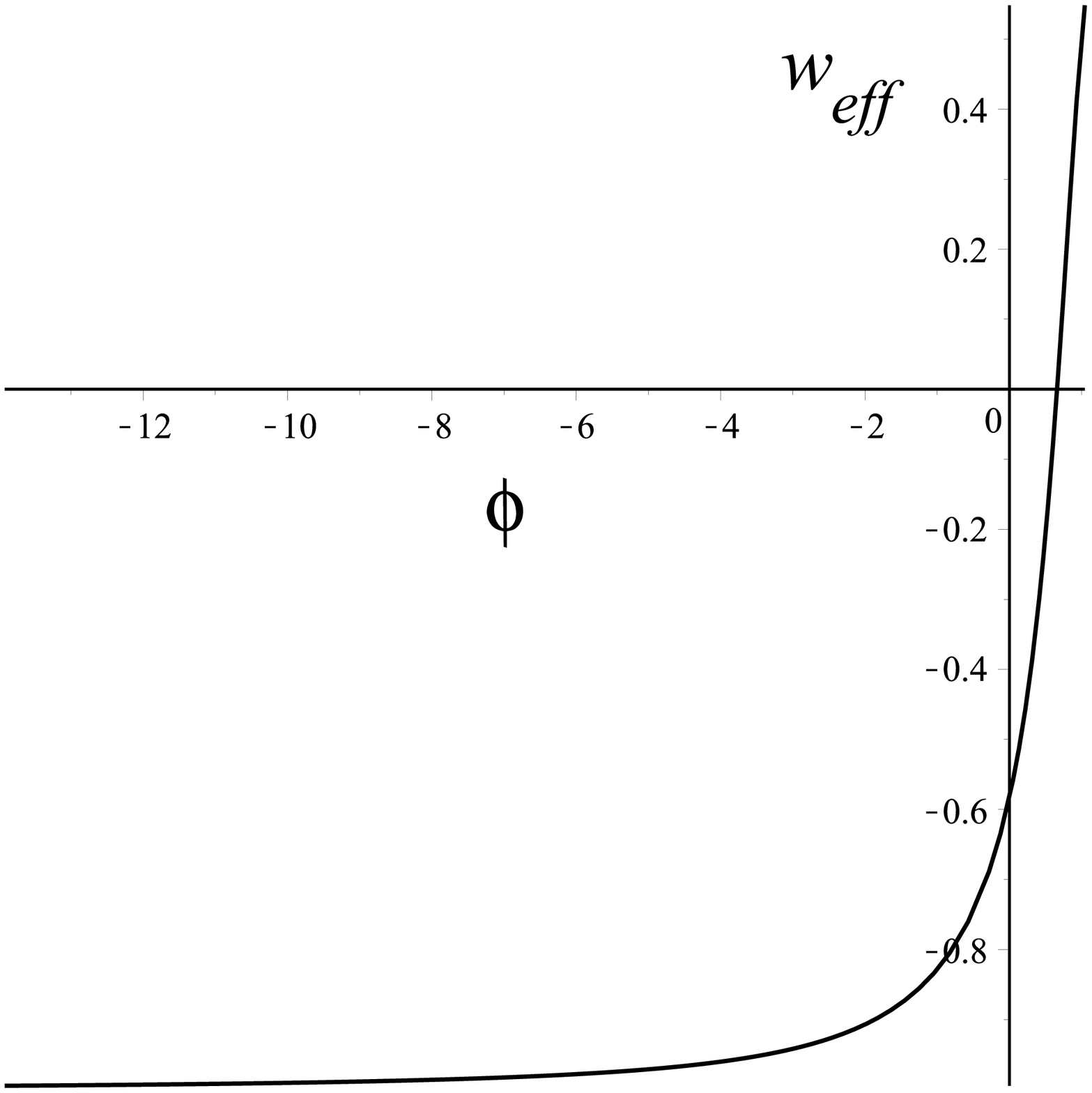}}
\caption{
The model equations (\ref{eqsNphi}) are numerically integrated for the choice of parameters $w_{\rm (in)}=-1$, $g=-4$ and $\alpha=2$. 
The behavior of the potential $V$ (in units of $\rho_{\rm (crit),0}$), the slow-roll parameters $\epsilon$, $\eta$ and $\Xi$, the number of $e-$folds $N$ during inflation and the ratio $p/\rho\equiv w_{\rm eff}$ between energy density and pressure as functions of $\phi$ (in units of $M_{\rm pl}$) is shown in panels (a) to (d), respectively. 
}
\label{fig:3}
\end{figure}

\begin{figure}
\begin{center}
\includegraphics[scale=0.4]{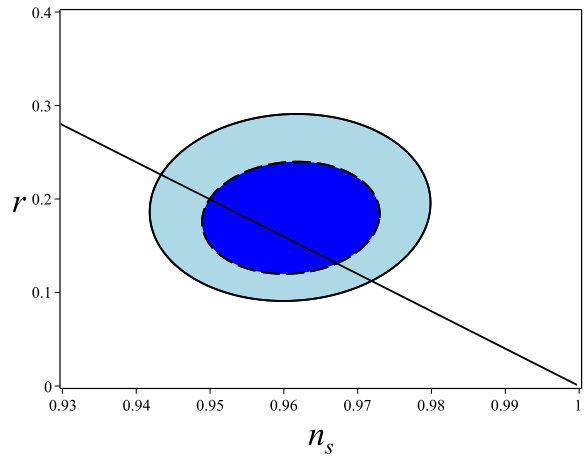}
\end{center}
\caption{
The contours of the ratio of tensor to scalar perturbations $r$ versus the scalar spectral index $n_s$ are {\it qualitatively} drawn from Planck+WP+highL+BICEP2 data \cite{bicep2}.
The straight line shows how $r$ changes as a function of $n_s$ for a SC model with the same choice of parameters as in Fig. \ref{fig:3}.
Noticeably, the same behavior is kept for a wide range of SC parameters $\alpha$ and $g$ that we have explored.
}
\label{fig:4}
\end{figure}

Let us turn to the general equations (\ref{eqsNphi}). 
An example of numerical integration is shown in Fig. \ref{fig:3} for the choice of SC parameters $w_{\rm (in)}=-1$, $g=-4$ and $\alpha=2$.
Fig. \ref{fig:3} (a) shows the behavior of the potential $V$ as a function of $\phi$.
It starts from a large value and monotonically decreases towards a minimum non-zero value, typical of hybrid models \cite{dodelson}.
In Fig. \ref{fig:3} (b), we plot the slow-roll parameters associated with the SC potential. The slow-roll regime ends due to $\epsilon$, which first approaches unity.  
Fig. \ref{fig:3} (c) shows the number of $e-$folds during the inflation matching the interval $N\approx[0,60]$ commonly assumed in inflationary models.
Finally, in Fig. \ref{fig:3} (d) we show the existence of a graceful exit mechanism from inflation by plotting as a function of $\phi$ the parameter $w_{\rm eff}$, i.e., the ratio between energy density and pressure.
With the above choice of SC parameters, we obtain $r\approx0.13$, $n_{s}\approx0.97$ and $\alpha_{s}\approx-5\times10^{-4}$, which are in agreement with observational data. 
Fig. \ref{fig:4} shows the agreement of our analysis with the joint constraints (at $1\sigma$ and  $2\sigma$ confidence level) on $n_s$ versus $r$ from current Planck+WP+highL+BICEP2 data (see Ref. \cite{bicep2} and references therein).


\section{Concluding remarks}

Summarizing, we have proposed a scalar field model for inflation based on the 
hydrodynamic analog of a Shan-Chen-like fluid.
A SC-like equation of state has been recently introduced in cosmology to represent the current distribution of dark energy, based on its property of evolving ordinary matter into a matter-energy component with an equation of state $p\approx-\rho$, 
through a phase transition mechanism.
In the context of inflation, we have considered a flat FRW universe filled by a canonical scalar field, with kinetic energy and potential related to   the SC energy density and pressure.
The evolution of the scalar field is thus completely determined by the SC dynamics, and we have analyzed in detail its properties in the slow-roll approximation.
Numerical inspection of the associated equations shows that simple choices of the free parameters of the SC model are consistent with current Planck, WMAP and BICEP2 data, i.e., the minimal viability requrement for any cosmological model. 
Furthermore, the equation of state undergoes a transition between $p/\rho<0$ (exotic matter) during inflation to $p/\rho>0$ (ordinary matter) at late times, thus 
providing also a graceful exit mechanism.
A more refined choice of parameters, as well as a suitable extension of the SC model presented here, are expected to match further available data.
Future directions of investigation will focus on the effects of
quantum fluctuations on the Shan-Chen cosmology and their
potential connections with the theory of hydrodynamic turbulence
\cite{ESS}.

\appendix

\section{Stability of scalar field inflationary models}

We briefly recall below the stability of scalar field inflationary models against scalar perturbations in the slow-roll approximation.
In the hydrodynamic approach one can define a speed of sound for the equivalent fluid as 
\beq
\label{csfluid}
c_s^2\equiv\frac{\partial p}{\partial\rho}
=\frac{\dot p}{\dot\rho}
=1+\frac{2V'}{3H\dot\phi}\,,
\eeq
where the definitions (\ref{rhoandp}) and the second equation of Eq. (\ref{ELeqns}) have been used.
Therefore, the slow-roll conditions (\ref{SLconds2}) imply $c_s^2\approx-1$, for every inflationary model. 
However, this does not necessarily imply the onset of instability for the fluid with respect to small-wavelength perturbations, as discussed in detail in Refs. \cite{dam_mukh,amend_mukh}. 
Following the standard theory of cosmological perturbations (see, e.g., Ref. \cite{mukhanov_phys_rept}), scalar perturbations of the background FRW metric due to small inhomogeneities of the scalar field $\phi(t,x^a)=\phi_0(t)+\delta\phi(t,x^a)$ evolve
according to:
\beq
\label{eqv}
\frac{d^2v}{d\eta^2}-{\tilde c_s}^2\nabla^2 v-\frac{1}{z}\frac{d^2z}{d\eta^2}v=0\,,
\eeq
for the canonical quantization variable $v=z\zeta$, defined in Ref. \cite{garr_mukh} (see Eq. (28) there and related discussion). 
In the above, $\eta$ denotes the conformal time, such that $dt=ad\eta$ and $(z,\zeta)$ are suitably defined perturbation functions.
As discussed in Ref. \cite{garr_mukh}, what is relevant for stability is the positiveness of the square of the \lq\lq effective speed of sound", appearing in 
front of the three-dimensional Laplacian in the perturbation 
equation (\ref{eqv}), defined as:
\beq
\label{cspert}
{\tilde c_s}^2\equiv\frac{p_X}{\rho_X}
=\frac{{\mathcal L}_X}{{\mathcal L}_X+2X{\mathcal L}_{XX}}\,,
\eeq
as per Eq. (\ref{rhoandpgen}). 
In the above, all quantities refer to their background values.

For a canonical scalar field, i.e., in the case considered in the present paper, we then obtain ${\tilde c_s}^2=1$, henceforth implying stability.
In the hydrodynamic representation one can relate $c_s^2$ to ${\tilde c}_s^2$ (see Eq. (10) of Ref. \cite{garr_mukh}), which can be written in the form
\beq
\frac{c_s^2-{\tilde c}_s^2}{1+{\tilde c}_s^2}=\frac{V'}{3H\dot\phi}\,
\eeq
which is consistent with $c_s^2\approx-1$ in the slow-roll regime.

\begin{acknowledgements}
DB acknowledges Prof. B. Mashhoon for useful discussions.
DG is supported by the Erasmus Mundus Joint Doctorate Program by Grant Number 2011-1640 from 
the EACEA of the European Commission.
\end{acknowledgements}

\end{document}